\documentclass[12pt,letterpaper]{article}

\usepackage{amsmath}
\usepackage{amsfonts}
\usepackage{amssymb}
\usepackage{graphicx}
\usepackage[numbers,sort&compress]{natbib}

\usepackage[hang,small,bf]{caption}

\DeclareRobustCommand{\SkipTocEntry}[4]{}

\def\beq{\begin{equation}}
\def\eeq{\end{equation}}
\def\bea{\begin{eqnarray}}
\def\eea{\end{eqnarray}}
\newcommand{\boldit}[1]{\textbf{\textit{#1}}}

\begin{document}

\vspace{5mm} \vspace{0.5cm}

\begin{center}

{\large  \bf Snowmass White Paper: Hamiltonian Truncation} 
	\\[1.0cm]

{ A. Liam Fitzpatrick$^{\rm 1}$, Emanuel Katz$^{\rm 1}$}
\\[.8cm]

%


{\small \textit{$^{\rm 1}$ Department of Physics, Boston University, Boston, MA 02215, USA}}

\end{center}

\vspace{1.5cm} \hrule \vspace{0.3cm}
{\small  \noindent \textbf{Abstract} \\[0.3cm]
\noindent 
Strongly-coupled Quantum Field Theories (QFTs) are ubiquitous in high energy physics and many-body physics, yet our ability to do precise computations in such systems remains limited.  Hamiltonian Truncation is a method for doing nonperturbative computations of real-time evolution in strongly coupled QFT in the continuum limit, and works by numerically solving the Schrodinger equation in a truncated subspace of the full Hilbert space.  Recent advances in understanding this method have opened the door to progress in a range of applications, from gauge theories in $d\ge 2$ dimensions to relativistic nonequilibrium physics.
 \vspace{0.5cm}  \hrule
\def\thefootnote{\arabic{footnote}}
\setcounter{footnote}{0}

\vspace{1.0cm}





\newpage

\section{Motivation}

\subsection{Limitations of the most common approaches to QFT}

Quantum Field Theory (QFT) has been very effective as a framework both in providing the language for the Standard Model (SM) as well as helping us parametrize possible Beyond-the-Standard-Model (BSM) scenarios.  For the most part, by QFT one usually means either the Perturbation Theory (PT) formulation, or the Lattice Field Theory (LFT) formulation.  Both techniques have been enormously successful, partly because Nature has been kind to us.  In particular, we have been very lucky that the SM is weakly coupled in the ultraviolet (UV) and that the only sector that is strongly coupled in the infrared (IR), Quantum Chromodynamics (QCD), has a lattice formulation (which for instance requires the empirical fact that the $\theta_{\rm QCD}$ parameter is negligible).  

Nevertheless, it is important to note that both PT and LFT have limitations and inefficiencies.  PT is an incomplete formulation, and clearly fails to describe physics when the coupling becomes strong.  However, it is also of note that PT fails even for so-called ``weakly coupled" theories in processes with a large multiplicity of particles, and so even at weak coupling some important phenomena are beyond its reach.  
For instance, at high energies  most  states of weakly coupled QFTs are expected to exhibit chaotic properties. In contrast to PT, LFT is a complete nonperturbative formulation of QFT, but as a universal approach to QFT it faces its own challenges.  Perhaps the most direct, is that a lattice is in fact a dynamical system in its own right, with potentially undesired physical consequences. Indeed, LFT is rather like a medium, with QFT degrees of freedom only emergent in the IR, typically when lattice parameters are tuned appropriately.  Phrased in this way, obtaining a particular desired QFT in the IR is not necessarily guaranteed, and can involve subtle lattice dynamics.  This appears to be the case for chiral gauge theories, like the SM.
It is noteworthy that despite the fact that LFT is a rather mature field with beautiful constructions (such as domain wall fermions for maintaining chiral symmetry) and
decades of development, it has thus far been unable to convincingly yield chiral gauge theories.  This is true even in the much simpler setting of 2d chiral gauge theories.  Other theories with non-trivial space-time symmetries, like supersymmetry, have also been difficult to realize using LFT.  Moreover, as is well known, even if in principle LFT dynamics manifest a certain QFT in the IR, practical considerations of Monte Carlo methods further require that the weight associated with field configuration on the lattice must be positive definite (i.e. cannot fluctuate in sign or have a phase).  Again, though lattices have been used extensively both in the High Energy Theory (HET) and the Condensed Matter Theory (CMT) communities for decades, a robust solution to this ``sign-problem" has not yet emerged.
It thus seems entirely plausible that the SM as well as many other interesting QFTs useful for BSM physics will not have practical LFT formulations anytime soon.

In addition, even for cases where an LFT formulation of a QFT is known, there are certain dynamical quantities that are difficult to extract directly due to the fact that the LFT formulation is typically Euclidean.  For example, a QCD observable like $e^+e^- \rightarrow hadrons$ is most directly related to the Wightman current two-point correlator as a function of real time, i.e.~in Lorentzian signature.  Similarly, obtaining information about bound-state wavefunctions, like determining the PDF of the proton is challenging to do directly.  Finally, there is no currently available LFT approach in circumstances when one is interested in real-time dynamics in an environment with a large multiplicity of particles, such as at finite temperature or chemical potential, (e.g.~for heavy ion collisions, or in the early universe, or with high particle multiplicity events in high energy scattering of scalars).  

\section{Recent Developments and Future Goals}


Hamiltonian Truncation is a framework for doing nonperturbative computations in quantum systems that works rather differently from PT or LFT.  In Hamiltonian truncation one attempts to approximate the eigenstates of the Hamiltonian by truncating the full Hilbert space of the theory to a well-chosen finite-dimensional subspace.
The resulting Hamiltonian in the truncated space is then a finite size matrix which can be numerically diagonalized.  Therefore, this approach uses only the degrees of freedom of the QFT itself, and can in many cases preserve the symmetries enjoyed by the untruncated theory.
Typically, the Hamiltonian is taken to be $H=H_0 +V$, where $H_0$ is integrable, with known eigenstates, while $V$ consists of a relevant local operator.  In the future, however, one could envision studying cases where the Hamiltonian, $H_0$, for the undeformed theory is not integrable, but is a CFT that can be numerically determined using the conformal bootstrap.  Unlike LFT, the QFT is realized directly, without describing it as the IR fluctuations of some other quantum system. 

QFT Hamiltonian truncation was begun by Brooks and Frautschi in 1984 \cite{brooks1984scalars},  but reached a wider audience in 1990 with  seminal papers by Yurov and Zamolodchikov \cite{yurov1990truncated,Yurov:1990kv,Lassig:1990xy}.  Their approach is known as the Truncated Conformal Space Approach (TCSA).  Here $H_0$ is the CFT Hamiltonian and the truncation is a cutoff on the energy of CFT states on the sphere (in the context of radial quantization).  This approach was quickly applied to many other RG flows in 2d relativistic field theories, including all deformations of the 2d Ising model and other minimal models, as well as the Sine-Gordon and Wess-Zumino-Witten models; for a recent thorough review, see \cite{james2018non}.   It was also later generalized to $d>2$ \cite{Hogervorst:2014rta}.  An appealing aspect of TCSA is that it was formulated in terms of CFT states, and therefore did not require a theory to have a Lagrangian description.  A complementary approach also emerged in the 1990s, named Discrete Light-cone Quantization (DLCQ) \cite{Pauli:1985ps}.  This was instead a Fock space method formulated in the infinite momentum or light-cone (LC) limit of a 2d theory.  Here $H_0$ is the free particle lightcone Hamiltonian and the truncation is on the total units of lightcone momentum carried by the particles.  DLCQ was mainly applied to 2d gauge theories and supersymmetric (SUSY) theories, both of which benefit from LC quantization's ability to preserve gauge invariance and part of the supersymmetry.  A more recent truncation scheme is Lightcone Conformal Truncation (LCT) \cite{Katz:2013qua,Chabysheva:2014rra,Katz:2016hxp, Anand:2020gnn}.  Here $H_0$ is the lightcone Hamiltonian of the CFT in Minkowski space and the truncation is on the scaling dimension of primary operators in the CFT basis.   In early work,  the above methods mostly focused on obtaining the energy spectrum for different theories in 2d.  However, the full set of observables and theories that can be studied with Hamiltonian Truncation is much richer.  In the following, we describe some of the main developments of this method in recent years, as well as  possible major applications in the future, and key goals for improving its efficiency and breadth.

\subsection{Gauge Theories}

One of the most important goals for Hamiltonian Truncation is to be a versatile tool for studying gauge theories at strong coupling.  The major challenge with studying gauge theories is that an energy or momentum cutoff in the Fock space framework violates gauge invariance.  In $d=2$, where gauge fields have no degrees of freedom, this problem can be solved by using lightcone quantization methods.  Consequently,  there have been a large number of nonperturbative studies of 2d nonabelian gauge theories coupled to matter fields.  These 2d methods are understood quite well at this point and it is known how to handle an arbitrary gauge group with scalar and matter fields in arbitrary representations \cite{Bhanot:1993xp, Demeterfi:1993rs,Gopakumar:2012gd,Anand:2020gnn,Dempsey:2021xpf}; for instance, they have produced important results on the phase structure of 2d QCD with adjoint matter. Ideas for preserving SUSY are also well studied 
\cite{Matsumura:1995kw,Hashimoto:1995jd,Antonuccio:1998jg,Lunin:1999ib,Hiller:2000nf,Fitzpatrick:2019cif}.  The space of Lagrangian theories containing scalars, fermions, and gauge fields is vast and fascinating, and provides an extremely rich set of toy models with which to explore strong dynamics and non-trivial RG-flows.  The chiral limit of non-abelian theories, in addition, is dual in the IR to deformations of many WZW and minimal models.  It was challenging to reach this chiral limit with earlier truncation methods; however, with new LCT technology the chiral frontier is now accessible, yielding precise results (see for example Fig.~\ref{fig:rhoTmm}, {\it right})  \cite{Anand:2021qnd}.   Another case which can be attempted with existing technology is that of 2d QED, coupled to scalars or fermions, with a nonzero $\theta$ parameter.  This example is expected to have an interesting phase structure, but is difficult to analyze using Monte Carlo due to a sign problem when $\theta \ne 0$.

\subsubsection*{Future Goals}

For $d>2$, however, new ideas must be implemented to maintain gauge invariance.  In principle, in $d=4$ the issue with gauge invariance can be resolved by making use of ideas from Conformal Field Theory.   
In particular, many interesting asymptotically free gauge theories can be embedded into a weakly coupled CFT by adding vector-like matter.  For instance, by adding more flavors, QCD can be made conformal and also weakly coupled (e.g. at $N_f=15$, $\alpha_s^* \approx 0.15$, while at $N_f=16$, $\alpha_s^* \approx 0.04$), known as a Banks-Zaks-Caswell fixed point.  One can then get back regular confining QCD simply by taking the weakly coupled CFT and deforming it by a relevant deformation -- namely, a mass for all the extra flavors that were added, $V = \int d^3 x ~ m \sum_i \bar{\psi}_i \psi_i$.  In this context, $H_0$ is just the CFT Hamiltonian for the fixed point, and the matrix elements of the deformation, $\langle {\cal O}_i | V | {\cal O}_j \rangle$ are in essence completely determined by the CFT data (i.e. the dimensions and OPE coefficients of the mass operator with all other operators) and by conformal kinematic functions which determine CFT three-point functions.  This data can be computed in perturbation theory or possibly by the conformal bootstrap in the future.
The CFT states are all gauge invariant (being related to local gauge invariant operators), and the relevant mass operator in $V$  is also local and gauge invariant.  Thus, the entire construction only uses gauge invariant data, and there is no danger that a UV-cutoff will spoil gauge invariance. 
In principle, TCSA and the LCT method, both of which are CFT deformations approaches, can make use of this construction.  Note, that this scheme can realize asymptotically free chiral gauge theories as well, by adding vector-like matter to make such theories conformal.

Implementing the above construction of 4d QCD  would be a major undertaking, however. The practical issue currently is that neither the CFT data nor the kinematic CFT functions in 4d (needed for LCT) have yet been computed.  It is thus more prudent to start with a UV fixed point theory about which more CFT data is known.  ${\cal N}=4$ SYM at large $N$ is a natural choice  because of integrability. It can be deformed by mass terms that preserve some but not all of the supersymmetry, providing additional control over UV divergences.  Still, much work is needed even here, involving a collection of expertise.  The kinematic CFT functions need to be parameterized properly and the specific CFT data needs to be extracted efficiently from integrability technology.  Both are significant projects that can be attempted in the immediate future, but must be completed before truncation work can begin.

Finally, gauge theories in $d=3$ are likely to be intermediate in difficulty between the $d=2$  and $d=4$ cases.  Because of their super-renormalizability, the breaking of gauge invariance by an energy cutoff may still be a problem that can be resolved by adding appropriate counterterms.  Recent advances \cite{Anand:2020qnp, Elias-Miro:2020qwz} in understanding how to handle counterterms in $d=3$ provide a concrete set of strategies that may be applied to 3d gauge theories.  A natural target is 3d QED, whose phase structure as a function of the number of flavors, $N_f$, remains an open question.  

\subsection{Correlators and Scattering Amplitudes from Eigenstates}

One of the main advantages of Hamiltonian Truncation is that by diagonalizing the Hamiltonian, one obtains the eigenvalue spectrum as well as the eigenvectors of the theory at strong coupling.  Together, these represent an enormous amount of information about the dynamics of the theory, allowing one to compute real-time evolution of any initial state.  Actually, one can compute both Lorentzian and Euclidean correlators, and Euclidean correlators of local operators in the vacuum, which are the correlators naturally computed in LFT, are straightforward to obtain from the eigenvector data.  For instance, the two-point function of a local operator ${\cal O}$ in the vacuum can be obtained through the evaluation of the spectral function $\rho_{\cal O}$ for ${\cal O}$, as a sum over states $\rho_{\cal O}(s) \sim \sum_j |\langle {\cal O}(0) |  j \rangle |^2 \delta(s-s_j)$.  The Euclidean two-point function is then the Euclidean free  propagator weighted by the spectral function. But now one can also use the spectral function to compute correlators in the Lorentzian regime as well, or indeed under any complex rotation of the momenta or positions! This procedure works much better than one might expect.  In particular, truncation discretizes the spectrum, so the spectral function becomes a sum over unphysical delta functions even when it should be a smooth function.  Nevertheless, methods that take advantage of unitarity and analyticity have been developed to resolve this issue \cite{Anand:2020gnn,Chen:2021bmm} and accurately restore the smooth underlying spectral functions.  In Fig.~\ref{fig:rhoTmm} ({\it right}), we show these smoothing methods applied to the Zamolodchikov $C$-function (the integral of the $T_{--}$ spectral function) in 2d QCD with $N_c=3$ and a very light quark in the fundamental.  This theory is strongly coupled in the IR, where it is expected to be dual to the Sine-Gordon model, allowing a precise test of the smoothed $C$-function from truncation.  Remarkably, even in this Lorentzian regime, Hamiltonian Truncation achieves better than 1\% accuracy using a basis of only 77 states.
These methods can likely be understood more thoroughly and improved. 

\subsubsection*{Future Goals}

Scattering amplitudes and multi-particle form factors are fundamental in QFT.
The difficulty in extracting them from the eigenstate data is that multiparticle asymptotic states are not eigenstates of the Hamiltonian.  The standard approach to extracting scattering amplitudes is the LSZ procedure, and developing LSZ applied to Hamiltonian truncation is likely to be a crucial area for future development.  There is also the potential for new, more indirect techniques for obtaining scattering amplitudes and form factors. Indeed, in $d=2$,   methods have been used to obtain scattering amplitudes from truncation data using analyticity and crossing symmetry rather than LSZ reduction \cite{james2018non, Chen:2021bmm, Gabai:2019ryw}.  In Fig.~\ref{fig:Amplitudes} ({\it right}), we show a result from  \cite{Gabai:2019ryw} for the elastic  S-matrix of the nonintegrable 2d Ising model with both a spin $\sigma$ and energy $\epsilon$ deformation, extracted indirectly from the eigenvector data of Hamiltonian Truncation calculations using Luscher's method \cite{Luscher:1985dn,Luscher:1986pf}; the presence of an unstable resonance shows up as a zero since the region plotted represents the physical (first) sheet in Mandelstam $s$.  We also show (Fig.~1,{\it left}) a result from \cite{Chen:2021bmm} for the elastic S-matrix in 2d $\phi^4$ theory for couplings ranging from the free theory up to the fixed point, extracted from Hamiltonian Truncation calculations;  at the fixed point, the result  matches the prediction of the free 2d Ising model.  In both of these results,  Hamiltonian Truncation data was combined with S-matrix bootstrap methods using analyticity and unitarity assumptions.  

\begin{figure}[t!]
\begin{center}
\includegraphics[width=0.55\textwidth]{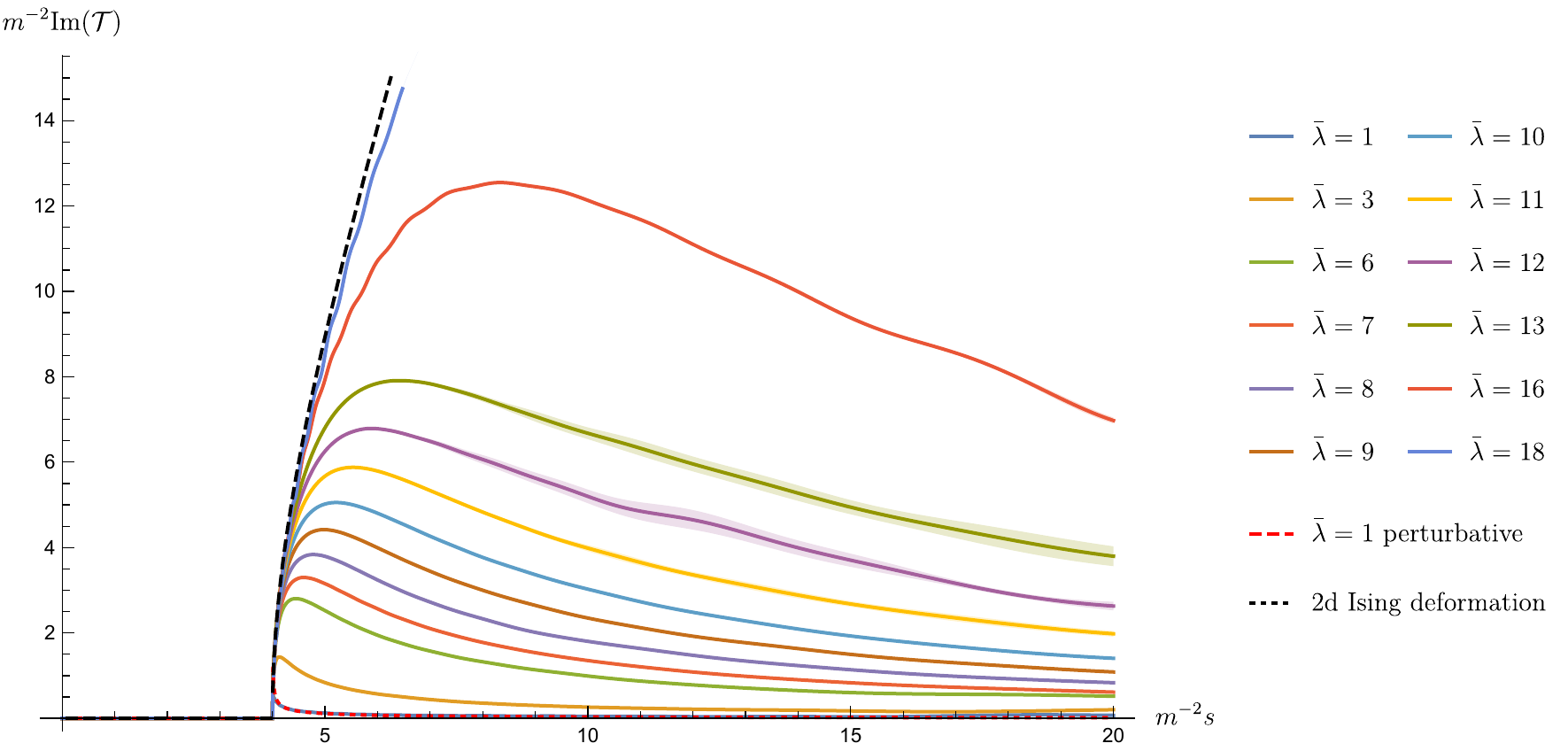}
\includegraphics[width=0.35\textwidth]{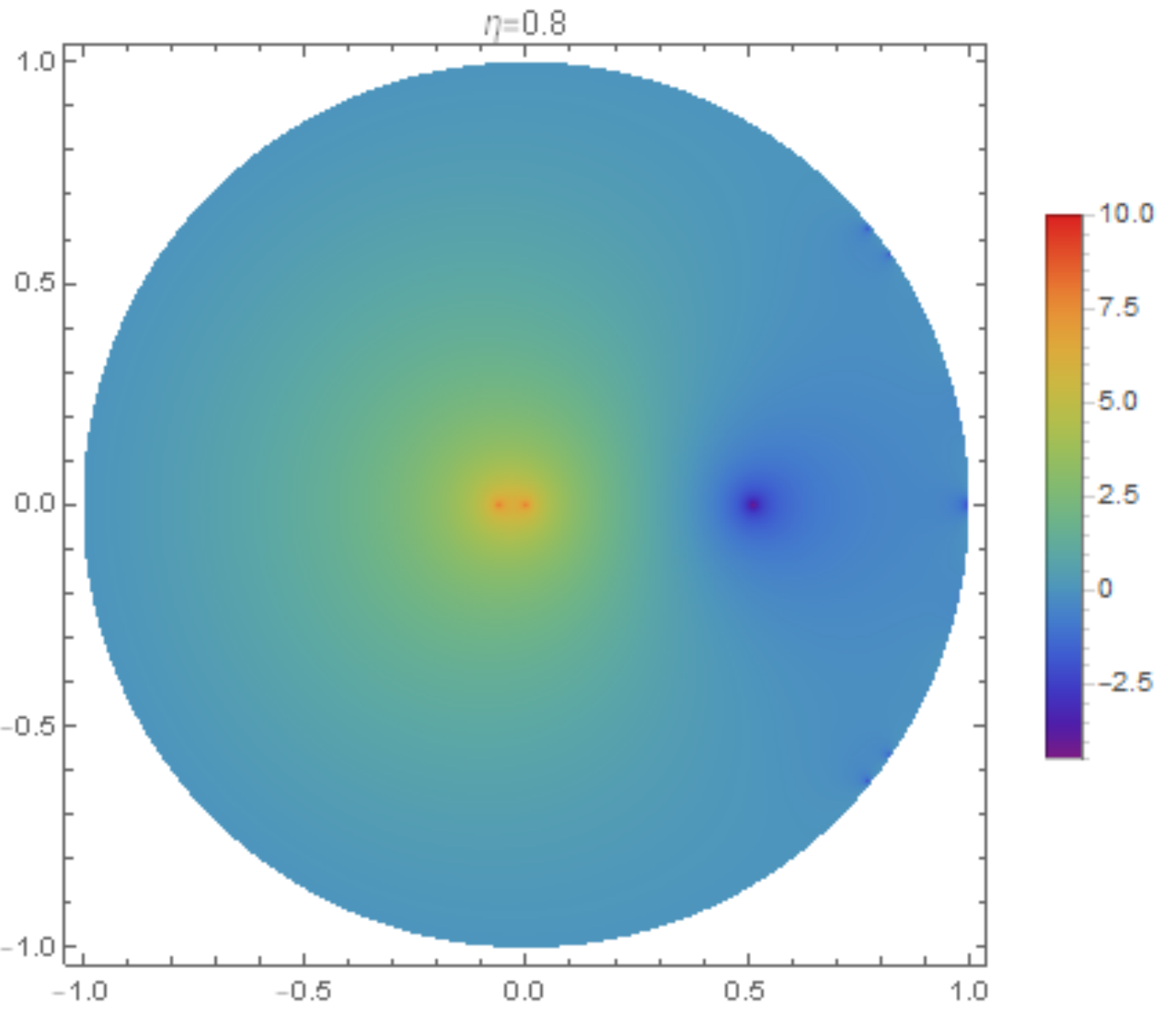}
\caption{Examples of Scattering Amplitudes in 2d theories calculated using Hamiltonian Truncation.  \boldit{Left:} (from \cite{Chen:2021bmm}) the elastic scattering amplitude in 2d $\phi^4$ theory, for couplings $\bar{\lambda}$ ranging from the free theory up to near the fixed point, where the amplitude matches the prediction of the free 2d Ising model.  \boldit{Right:} A plot (from  \cite{Gabai:2019ryw}) of $\log |S(z)|$, where $S(z)$ is the  elastic scattering amplitude in the 2d Ising model with both a spin $\sigma$ and energy $\epsilon$ deformation, analytically continued to the complex plane in the variable $z$, where $( \frac{1+z}{1-z})^2 = \frac{s (s-4m^2)}{3m^4}$, which maps the first sheet in Mandelstam $s$ into the unit disk. }
\label{fig:Amplitudes}
\end{center}
\end{figure}

There is also an important connection between scattering amplitudes and spectral densities, when a strongly coupled theory is perturbatively coupled to an external probe: the spectral densities themselves directly provide the scattering amplitudes of the weakly coupled probe particles.   This is the case for instance with using electromagnetic scattering as a probe of QCD.
In this way, the 2d QCD spectral density shown in Fig.~\ref{fig:rhoTmm} ({\it left}) can be directly interpreted as a scattering cross-section similar to that of $e^+ e^- \rightarrow$ hadrons in 4d QCD.  So, while much work remains to be done, 
in principle there is a path towards computing $\sigma(e^+ e^- \rightarrow$ hadrons) in 4d QCD with Hamiltonian truncation, making it an important target for the future.

\begin{figure}[t!]
\begin{center}
\includegraphics[width=0.5\textwidth]{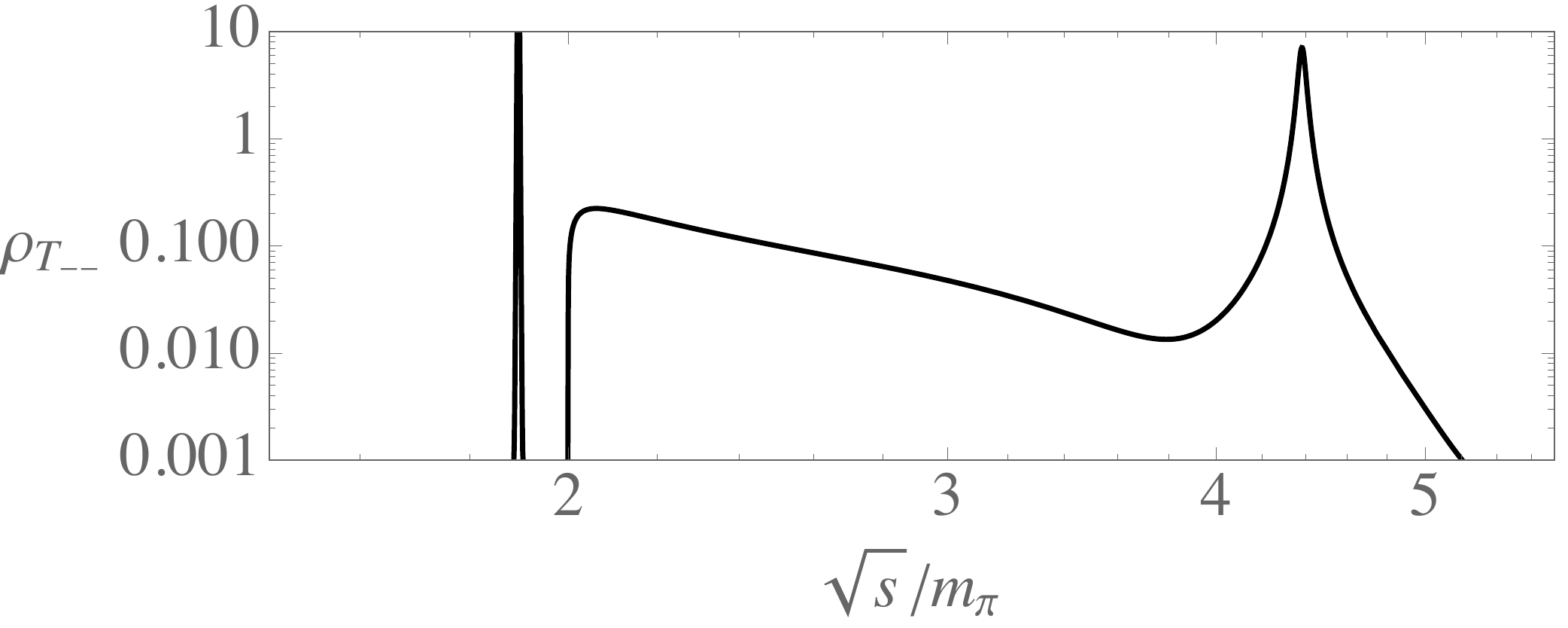}
\includegraphics[width=0.4\textwidth]{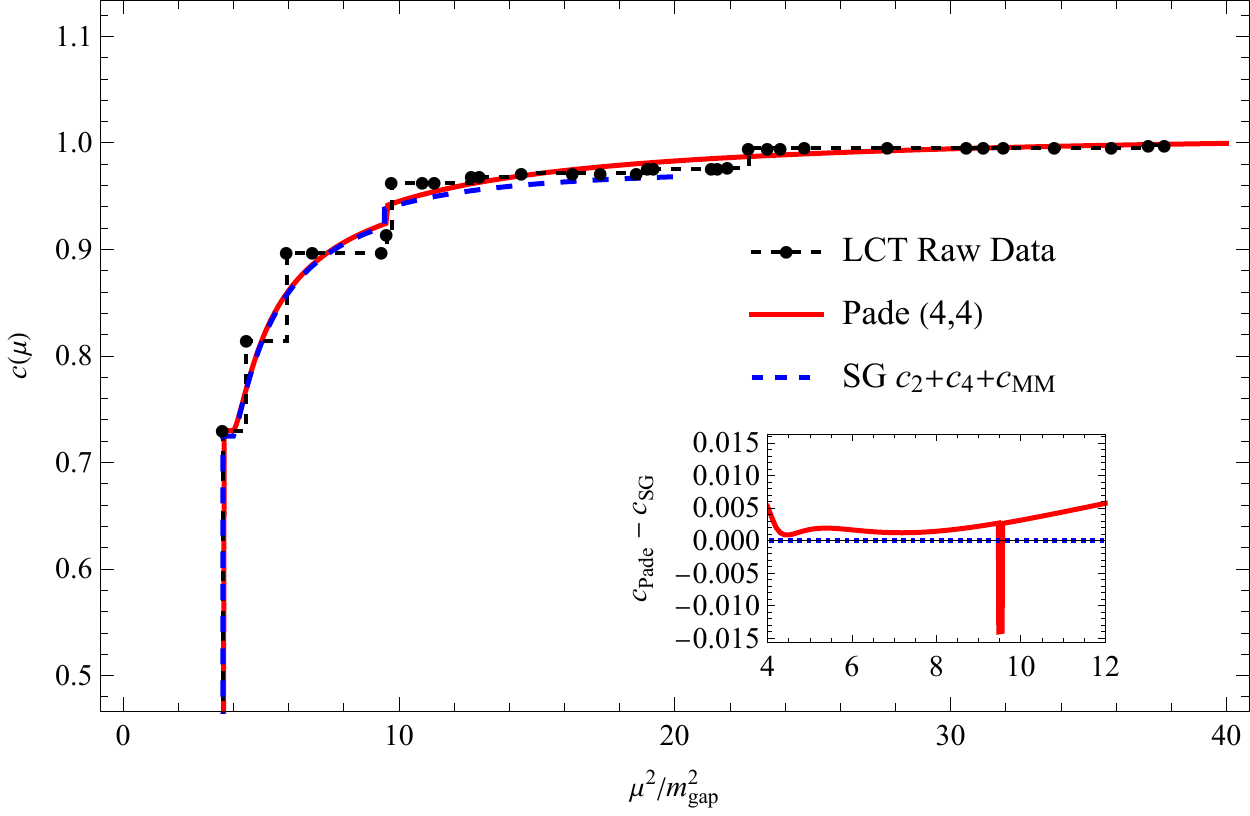}
\caption{\boldit{Left:} A plot of a Hamiltonian Truncation computation of the spectral density for the stress tensor $T_{\mu\nu}$ as a function of center-of-mass energy $\sqrt{s}$, in 2d QCD with $N_c=3$ colors and a single massive fundamental quark. The theory contains a light scalar `pion' meson with mass $m_\pi$, and a stable bound state of two `pions'  at $\sqrt{s} \lesssim 2 m_\pi$. 
An unstable resonance is also visible at around $\sqrt{s} \sim 4.4 m_\pi$. 
\boldit{Right:} A plot (from \cite{Anand:2021qnd}) of the Zamolodchikov $C$-function (which is the integral of the stress tensor spectral density), for a very small value of the quark mass in which case the theory is expected to be dual to the Sine-Gordon model. {\it Black, dashed}: the raw $C$-function from truncation, which contains unphysical steps due to the discreteness of the truncated spectrum.  {\it Red, solid}: the improved $C$-function, after smoothing. 
{\it Blue, dashed}: the $C$-function in the Sine-Gordon model computed using integrability methods. The relative error between the Sine-Gordon $C$-function and the improved $C$-function from truncation is shown in the inset.}
\label{fig:rhoTmm}
\end{center}
\end{figure}

\subsection{Improved Truncation and Renormalization Procedures}

A generic property of truncation schemes is the rapid growth in the size of the space of states as the truncation parameter is increased.  Consequently, a concern is that obtaining results with the desired high level of precision could require an intractably large subspace in Hamiltonian Truncation. 

One strategy for overcoming this problem is  to find more efficient truncation subspaces  that lead to faster convergence.  For instance, in \cite{Vary:2021cbh}, the authors studied 2d $\phi^4$ theory using DLCQ, and with improved algorithms and computational resources the authors were able to take the truncation parameter $K=96$, corresponding to about 60 million states.  Such a large truncation made it possible to reliably establish that the truncation error in the critical coupling $\lambda_c$ converged to zero like $O(\frac{1}{\sqrt{K}})$ as $K\rightarrow \infty$.  By contrast, in the LCT basis, the truncation error converges like $O(\frac{1}{K})$; in fact, to good approximation, the critical coupling in LCT and DLCQ as a function of $K$ are related by $\lambda_c^{(\rm LCT)}(K) = \lambda_c^{(\rm DLCQ)}(K^2)$.\footnote{In 2d $\phi^4$ theory in both DLCQ and LCT, the number of states as a function of $K$ is the number $P(K)$ of integer partitions of $K$, which grows rapidly with $K$ like $P(K) \approx \frac{1}{4 K \sqrt{3}} e^{ \pi \sqrt{\frac{2}{3}K}}$; for instance, $P(20)= 627$, while $P(400)=6.7 \times 10^{18}$. } 
 So,  different choices can lead to vastly different rates of convergence as the size of the subspace is increased. Additional strategies can be employed to identify more efficient subspaces that keep only the most important states in the basis; see section VII of \cite{james2018non} for  examples.  

A second strategy is to systematically approximate the contributions of states above the truncation. Here, significant improvements have come from recent progress on how to formulate  low-energy effective descriptions for cutoffs in Hamiltonian frameworks \cite{Rychkov:2014eea,Elias-Miro:2017tup,Cohen:2021erm}. Such progress not only allows one to improve the rate of convergence by explicitly computing the leading corrections in the truncation parameter, but moreover provides a better understanding of the expected form and truncation-dependence of such corrections.  For example, the improvements of \cite{Elias-Miro:2017tup} yielded a very precise value of the critical coupling for 2d $\phi^4$, as well as various other data near criticality (see Fig.~\ref{fig:criticalspec}).  Also, quite importantly, understanding the effect of truncation cutoffs is crucial for handling UV divergences in a truncation framework.

In fact, the difficulty of renormalization with an energy cutoff is one of the most conceptually challenging issues facing the implementation of Hamiltonian Truncation in a wider range of theories.    A hard cutoff in the total energy of the system is a nonlocal condition, and relates distant regions of spacetime to each other. Consequently, the counterterms that are required to cancel divergences often cannot be represented as local interactions, and are much more difficult to understand \cite{perry1990light}.  However, there have been a number of significant breakthroughs in the past few years \cite{Elias-Miro:2020qwz, Anand:2020qnp, Rutter:2018aog,Hogervorst:2014rta,EliasMiro:2021aof} in the treatment of such counterterms. As a result, it has become possible to use Hamiltonian truncation in many super-renormalizable theories with UV divergences in $d>2$.  In particular, \cite{Elias-Miro:2020qwz, Anand:2020qnp} recently completed Hamiltonian truncation studies of strongly coupled 3d $\phi^4$ theory.  A highly nontrivial check that the method employed in \cite{Anand:2020qnp} is working is that  near the critical point, correlators of operators exhibit IR universality over a range of scales, seen by different UV operators, with the expected scaling exponents (shown in Fig.~\ref{fig:3dphi4}, {\it right}). Likewise, the method employed in \cite{Elias-Miro:2020qwz} passes the nontrivial test that the spectrum satisfies a strong-weak duality of the model (Fig.~\ref{fig:3dphi4}, {\it left}).

\begin{figure}[t!]
\begin{center}
\includegraphics[width=0.4\textwidth]{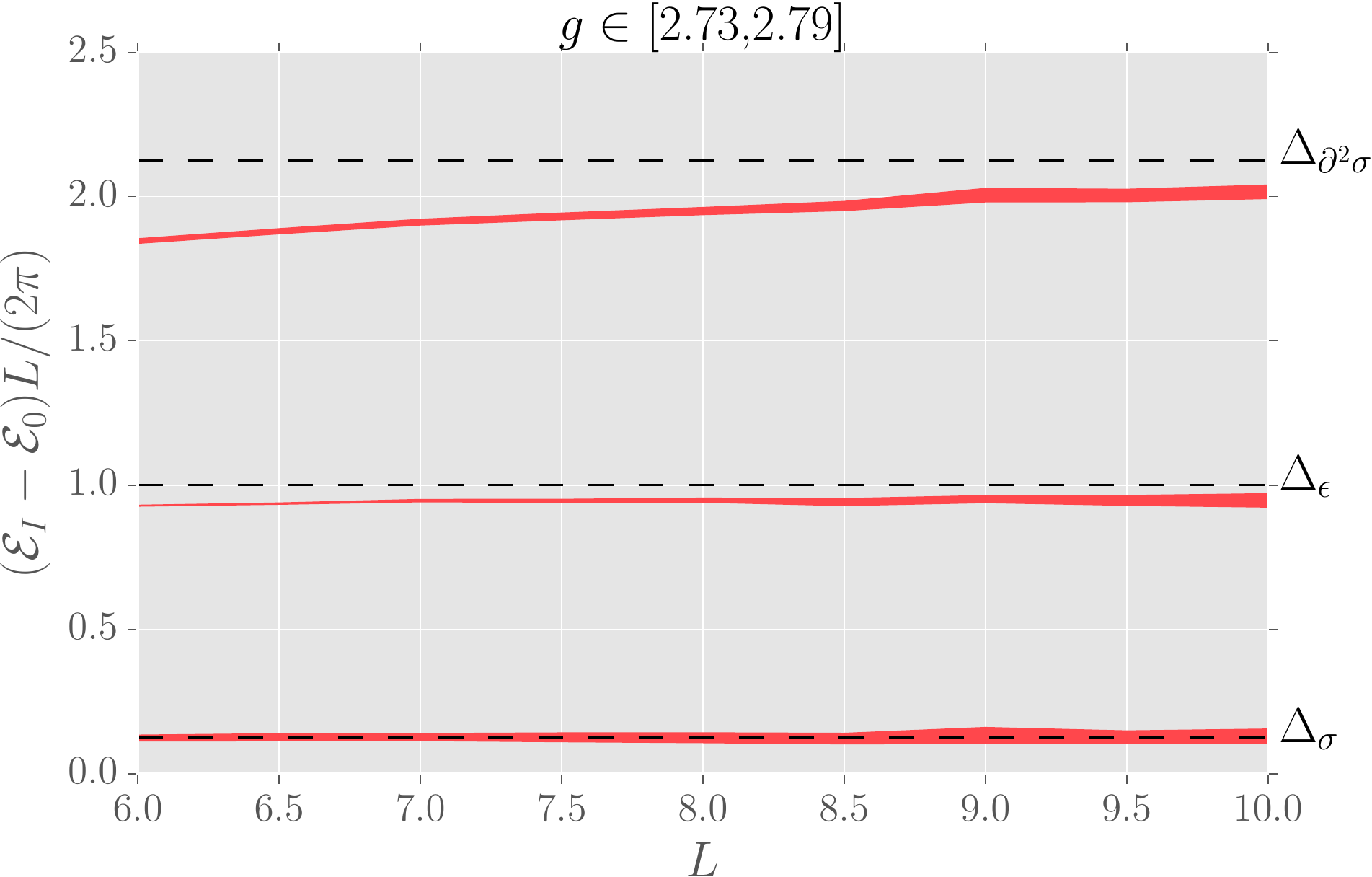}
\caption{Comparison of energy levels at the critical coupling for 2d $\phi^4$ with CFT predictions (from \cite{Elias-Miro:2017tup}).}
\label{fig:criticalspec}
\end{center}
\end{figure}

\begin{figure}[t!]
\begin{center}
\includegraphics[width=0.4\textwidth]{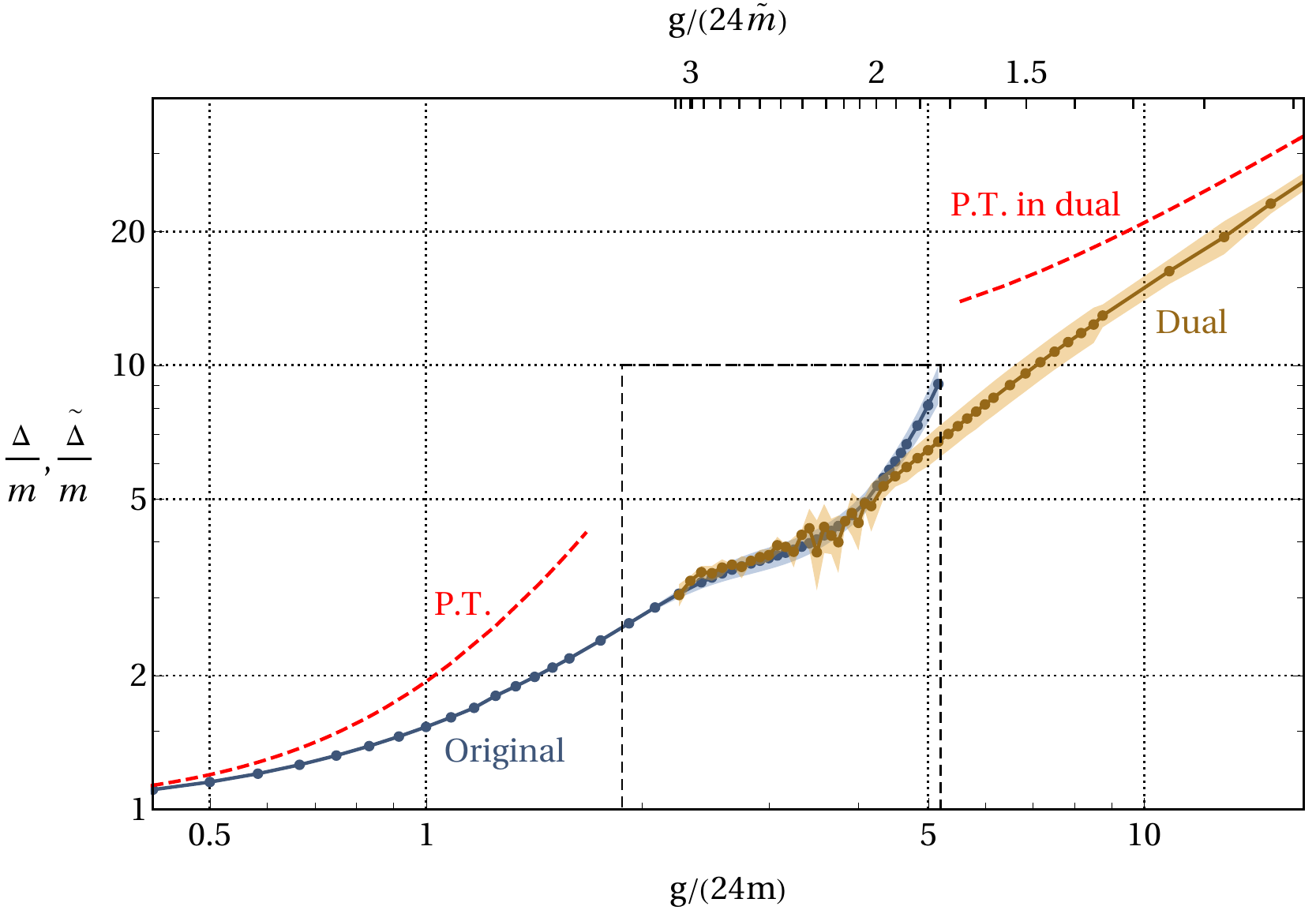}
\includegraphics[width=0.4\textwidth]{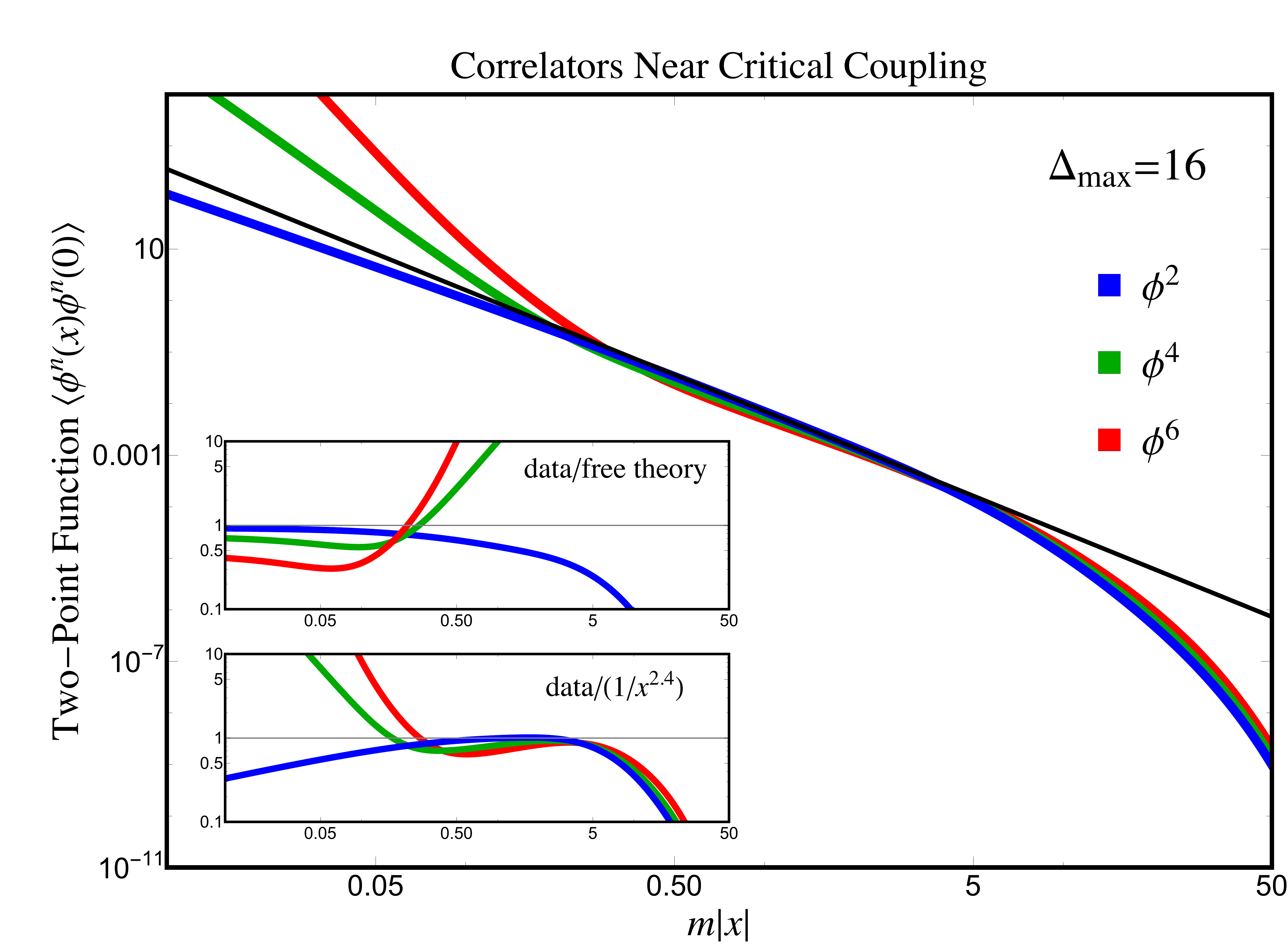}
\caption{Hamiltonian Truncation results for 3d $\phi^4$ theory at strong coupling.  \boldit{Left}: The mass spectrum from \cite{Elias-Miro:2020qwz} is explicitly seen to satisfy a strong-weak duality.  \boldit{Right:} the correlators of various operators from \cite{Anand:2020qnp} near the critical point obey universal IR scaling behavior, consistently with the known critical exponents of the 3d Ising model. }
\label{fig:3dphi4}
\end{center}
\end{figure}

\subsubsection*{Future Goals}

Improving the truncation procedure will  require developing more efficient methods for obtaining the Hamiltonian matrix elements themselves.  Finding these matrix elements is often a highly nontrivial task.  In TCSA and LCT, they are equivalent to OPE coefficients of primary operators in the CFT of the ultraviolet fixed point.  Even in the case where the UV fixed point is a free theory, these OPE coefficients are computationally expensive at very large truncations, due to the  complexity of high-dimension primary operators.  In fact, already, taking advantage of the conformal structure of the UV theory operators has been crucial to being able to compute these OPE coefficients efficiently, allowing for work with very large bases \cite{Anand:2020gnn,Anand:2019lkt}.  Moreover, in the LCT framework the Hamiltonian matrix elements involve kinematic factors, which are essentially the Fourier transforms of CFT three-point functions.  While these are known in closed form in 2d, in $d>2$ the required Fourier transforms are still not known in general.  

In addition, there is more work to be done on understanding counterterms in Hamiltonian Truncation frameworks. Even with recent improvements, the available procedures for dealing with  counterterms can be ad hoc at times, and further work is needed.  Moreover, they should be employed and tested in a wider range of theories, as at present most $d>2$ work has been limited to scalar field theories.  A natural extension is to a 3d Yukawa theory, as possibly the simplest theory with both scalars and fermions.

\subsection{Non-Equilibrium Physics}

Because Hamiltonian Truncation produces the nonperturbative eigenvalues and eigenvectors of a theory, it provides one of the only known methods for calculating real-time non-equilibrium dynamics in strongly coupled, relativistic systems. Various measures of quantum chaos can be calculated and compared with the predictions of Random Matrix Theory (RMT). 
Eigenstates can be used to directly construct canonical and microcanonical ensembles, or to test eigenstate thermalization, and nonequilibrium fluctuations around such backgrounds can then be computed from correlators of local operators \cite{James:2018qly,Robinson:2018wbx,rakovszky2016hamiltonian}.

Most quantities of this nature can already be computed in practice in simple 2d theories.  For instance, using  the distribution of eigenvalue spacings, one can observe the chaotic nature of 2d $\phi^4$ theory at both weak and strong couplings (see Fig.~\ref{fig:WD}, {\it left plot}), and more generally in deformations of rational theories \cite{Brandino:2010sv,Srdinsek:2020bpq}; one can also study the emergence of chaos through the statistics of eigenvector components.  From the density of eigenvalues as a function of energy, one can extract the entropy and other thermodynamic quantities. This includes quantities which characterize hydrodynamics in 2d, like the speed of sound, and the KPZ diffusion constant \cite{Delacretaz:2021ufg}.  Thus, truncation is unique in its ability to capture both thermalization and approaches to equilibrium (as expected of UV relativistic systems), followed by late time behavior indicative of chaos.  Indeed, other numerical approaches to nonequilbrium physics use lattices, which violate space-time symmetries and whose typical excited states are rather different from relativistic QFTs.  Calculating the Spectral Form Factor (SFF) exhibits the flexibility enjoyed by truncation methods (see Fig.~\ref{fig:WD}, {\it right plot}): It is a real-time observable whose early time behavior is determined by the physical density of states of the QFT, but whose late time behavior is controlled by universal chaotic dynamics as dictated by RMT.

\begin{figure}[t!]
\begin{center}
\includegraphics[width=0.4\textwidth]{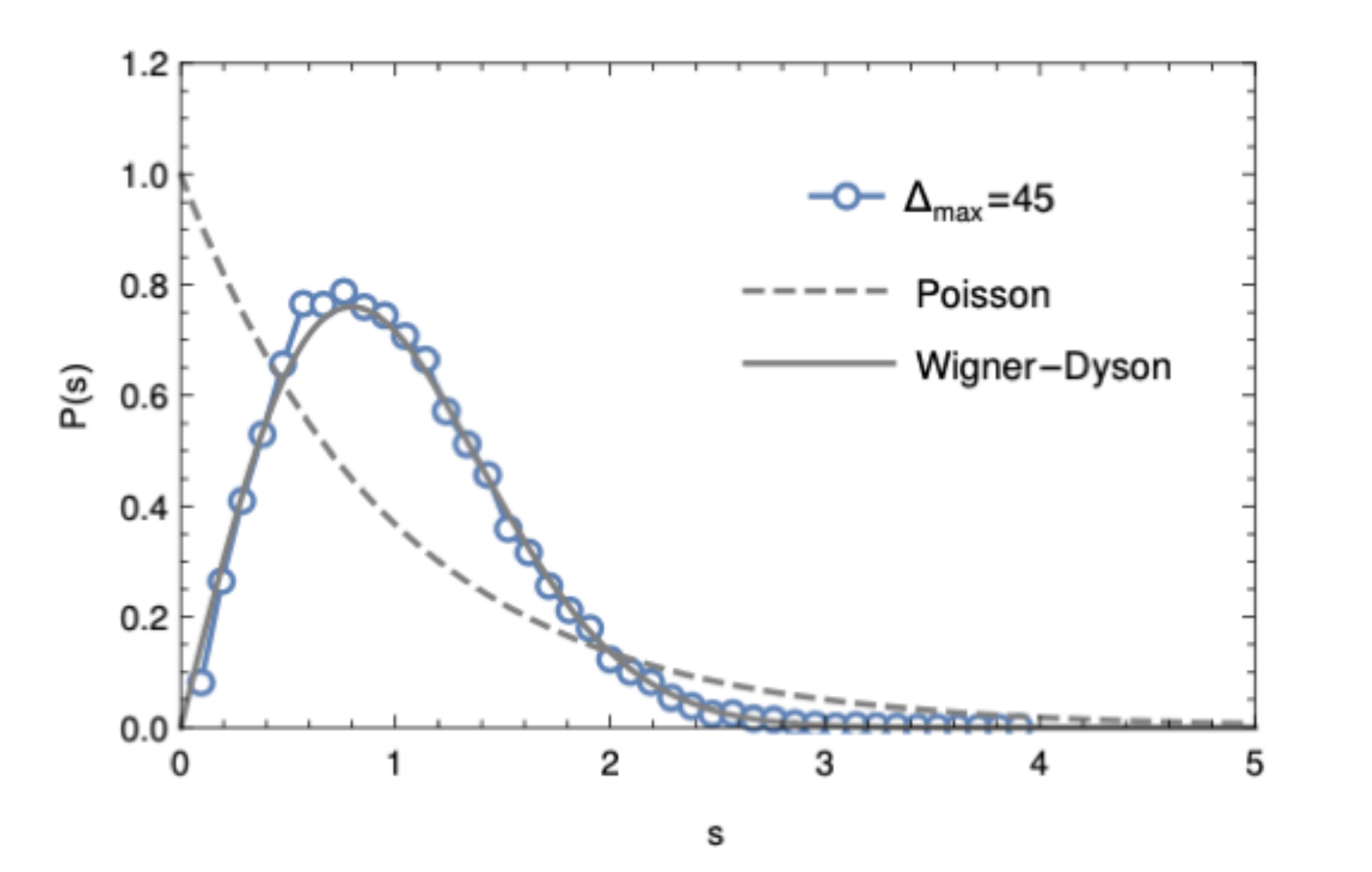}
\includegraphics[width=0.4\textwidth]{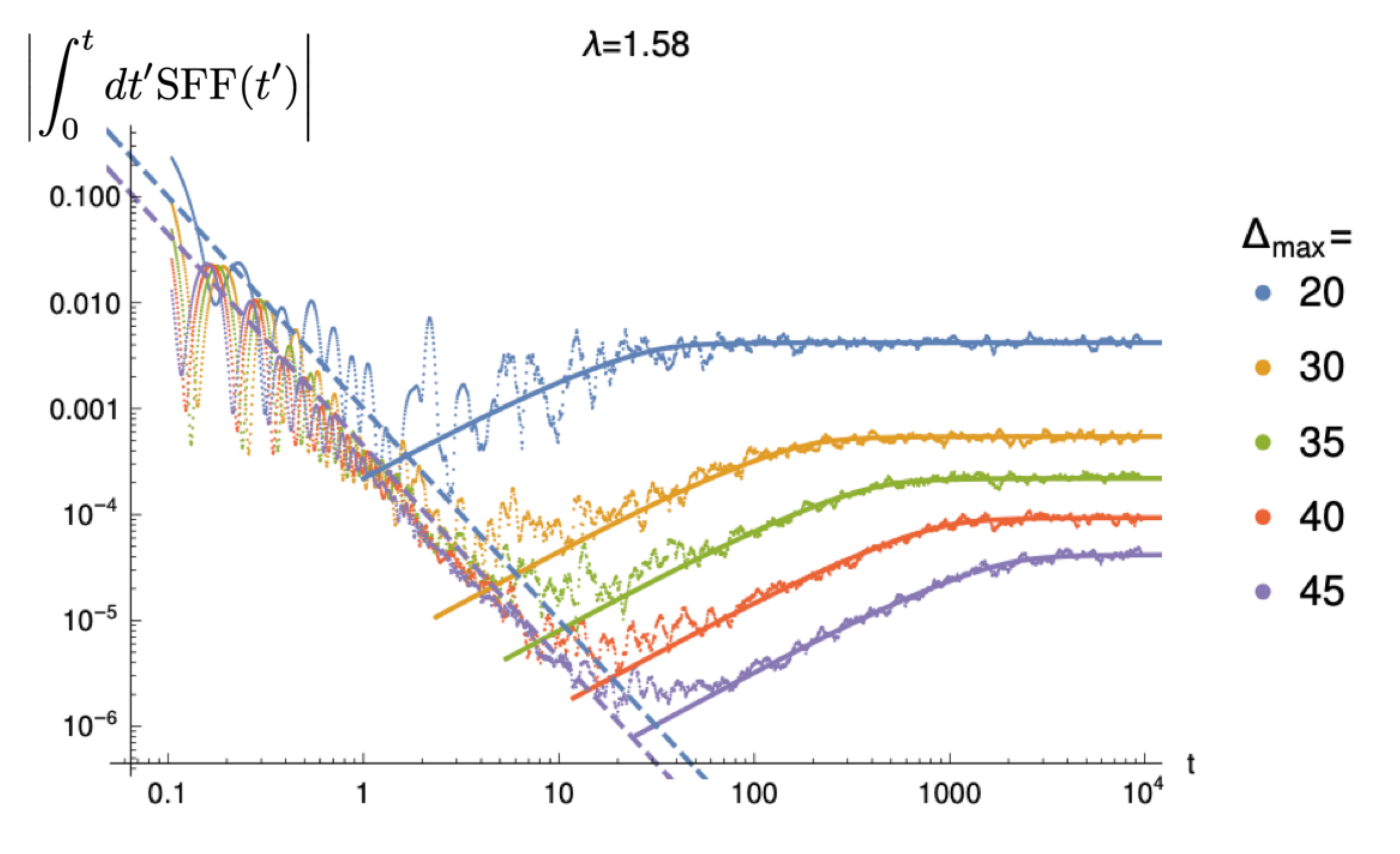}
\caption{\boldit{Left:} Distribution of eigenvalue spacings in 2d $\phi^4$ theory from Hamiltonian Truncation.  The distribution closely follows the Wigner-Dyson distribution predicted by RMT, rather than a Poisson distribution which is characteristic of integrable theories.  Similar results have been obtained in deformations of rational 2d CFTs in  \cite{Brandino:2010sv}. \boldit{Right:} The magnitude of the integrated Spectral Form Factor, SFF$(t) \equiv \sum_{i,j} e^{i(E_i - E_j) t }$ ({\it show as data points}) in strongly coupled 2d $\phi^4$ theory, for various values of the truncation parameter $\Delta_{\rm max}$,  compared against the late-time RMT predictions ({\it solid lines}) and the early-time prediction from the density of states ({\it dashed lines}). Plots from \cite{private}.}
\label{fig:WD}
\end{center}
\end{figure}

\subsubsection*{Future Goals}

Extending this approach holds the potential to address a wide range of questions about nonequilibrium physics in strongly coupled theories.  The study of relativistic hydrodynamics in particular presents an opportunity for Hamiltonian Truncation in continuum QFT, since the presence of a lattice in most other strongly coupled numeric approaches breaks the translation invariance that is essential for the long-lived hydrodynamic modes.  The inclusion of theories with conserved global symmetries would also allow one to study diffusion and the effect of chemical potentials.  In particular, an important goal is to apply Hamiltonian truncation to study transport in the 3d $O(2)$ model, which describes superfluids and many other systems at low energies. The wealth of data provided by the knowledge of eigenvectors in Hamiltonian Truncation offers many exciting opportunities for progress in the future.

\begin{center}
\subsection*{Acknowledgments}
\end{center}

We thank Joan Elias-Mir\'o, Robert Konik, and Slava Rychkov  for comments on an earlier draft. ALF and EK were supported in part by the US Department of Energy Office of Science under Award Number DE-SC0015845. 

\newpage

\bibliographystyle{JHEP}
\bibliography{refs}

\end{document}